\documentclass[aps,pra,floatfix,amsmath,amssymb,superscriptaddress]{revtex4}

\usepackage{latexsym}
\usepackage{epsfig}
\usepackage{dcolumn}% Align table columns on decimal point
\usepackage{amsmath}
\usepackage{amssymb}
\usepackage{graphicx}
\usepackage{times}
\usepackage[section]{placeins}
\usepackage{caption}

\usepackage{color}

\def\m{\mu}
\def\n{\nu}
\def\be{\begin{equation}}
\def\ee{\end{equation}}
\def\ba{\begin{eqnarray}}
\def\ea{\end{eqnarray}}
\def\la{\langle}
\def\ra{\rangle}

\def\m{\mu}
\def\n{\nu}

\def\h{\hskip 1cm}

\def\lo{\longrightarrow}

\begin{document}

	\title{Quantum Key Distribution with no Shared Reference Frame}
\vspace{2cm}
	\author{F. Rezazadeh}
	\affiliation{Department of Physics, Sharif University of Technology, Tehran, Iran}
	
	\author{A. Mani}
	\affiliation{Department of Engineering Science, College of Engineering, University of Tehran, Iran}
	
	\author{V. Karimipour}
	\affiliation{Department of Physics, Sharif University of Technology, Tehran, Iran}
	
\vspace{2cm}

	\begin{abstract}
		Any quantum communication task requires a common reference frame (i.e. phase, coordinate system). In particular Quantum Key Distribution requires different bases for preparation and measurements of states which are obviously based on the existence of a common frame of reference. Here we show how QKD can be achieved in the absence of any common frame of reference. We study the coordinate reference frame, where the two parties do not even share a single direction, but the method can be generalized to other general frames of reference, pertaining to other groups of transformations. 
		\end{abstract}
	
	%Our emphasis is entirely theoretical and do not propose a concrete experimentaassumes the existence of a shared reference frame between the two parties. Otherwise there will be no correlation between the states prepared and measured by the two legitimate players. Here we present a method which doesn't need any reference frame, nor even a single  direction  
%	to be shared between two remote labs.  In principle the method can be used in QKD schemes between earth- and satellite-based labs where coordinates systems are constantly changing, although its feasibility remains for the future.}   

	
	\date{\today}
	
	\maketitle
\vspace{0cm}

\section{Introduction}

Any quantum communication protocol requires that the parties involved do have a Shared Reference Frame (SRF). The absence of such a common reference frame and their relation is equivalent to the existence of an almost depolarizing  channel between the two parties. The reason is that any state $\rho_A$ sent by $A$ is seen by $B$ to have been transmitted through a channel
\be\label{basicc}
\rho_A\lo \rho_B=\int dg P(g)U(g)\rho_A U^{\dagger}(g),
\ee
where $g\in G$, and $G$ is the group of transformations connecting  the two reference frames, and $U(g)$ is a unitary (not necessarily irreducible) representation of the group and $P(g)$ represents the partial knowledge that we have on the misalignment of the two frames. The group measure is also left and right invariant, i.e.  $dg=d(gh)=d(hg)\ \ \forall h\in G$. When the state $\rho$ carries an irreducible representation of the group $G$, and $P(g)$ is uniform,  the state $\rho_B$ will be a completely mixed state $\rho_B\propto I$, due to the Shur's lemma.   The group $G$  depends on the variable which the state encodes, i.e. it can be $U(1)$ if the variable is a phase or $ SO(3)\sim SU(2)/Z_2$ if the variable is a direction in space. In practice we may align the two frames by other possibly classical means, but it is a theoretically interesting question to ask, how we can do  quantum communication in the complete absence of common reference frames. There has been intensive interest on the subject of reference frames \cite{aligning reference frame1,aligning reference frame2,aligning reference frame3,aligning reference frame4,communication without SRF1,communication without SRF2,marvian} and there has been reports on different protocols ranging from tests of Bell inequality in the absence of reference frames \cite{nonlocality without SRF1,nonlocality without SRF2,nonlocality without SRF3} to quantum key distribution when the two players only share a common direction \cite{RFI-QKD1,RFI-QKD2,RFI-QKD4}. Nevertheless the problem of Quantum Key Distribution, as one the most important quantum communication tasks has not been studied in its generality. This is the subject of the present  paper.\\ 
%\textcolor{red}{
%The  idea of Quantum Key Distribution (QKD) \cite{BB84, Ekert} was to base the security of a key distribution not on the hardness of certain mathematical problems but on the inherent impossibility of measuring incompatible observables of quantum systems. The former schemes, no matter how complex they are, do not have forward secrecy, that is they are  not secure against inevitable advances in our capability of solving mathematical problems in the future, either with classical or quantum computers, while the latter are unconditionally  secure since they are based on inherent laws of the natural world as conceived in quantum mechanics.  However once the original excitement of the basic idea is over, we are faced with practical challenges of how to implement the idea in real situations. A number of these challenges are imperfections in single-photon generators  and detectors , photon losses in long distance communications, the need for high-speed quantum random number generators and broadband entangled photon sources, for a recent review see 
%\cite{nature}. }\\
Our motivation stems from the unique features of this task, namely: the necessity of using different random bases for preparation of states on one side and different random bases for measurements on the other, both of which depend crucially on the existence of common reference frames. It is thus a pressing question that in the absence of any common reference frame, when the two players do not share even a single direction,  how such a protocol can be run.\\

In the following, although we emphasize on one type of reference frame, namely a coordinate system, it is fairly clear how our considerations can be generalized to other types of frames, whose transformations belong to a different group. 
The question we ask is how the two players (conventionally called Alice and Bob) can run a QKD protocol when they do not share any coordinate system nor even a single direction. A QKD scheme is essentially based on the notion that a state which has no coherence in one basis (i.e. $|z_+\ra$ in the case of spin 1/2 particles) is maximally coherent in another basis ($\frac{1}{\sqrt{2}}(|x_+\ra+|x_-\ra)$) \cite{BB84, Ekert, 6-state, cerf, karl, vk}.
 Therefore if measured in the wrong basis, such a state will produce completely random and hence uncorrelated results with those of the other player. However in the absence of frames of reference, not all  superpositions are well defined, as they are forbidden by superselection rules \cite{SSR1,SSR2}. Here we propose a scheme which solves this problem and explain it in the context of coordinate reference frames, when the degrees of freedom are those of spin 1/2 particles or photons. The basic idea can be generalized to other frames and their corresponding groups of transformations.  Essentially it is based on encoding the variable of interest in the fusion space of representations of the group, as we will discuss at the end of the paper.\\

Our emphasis here is mostly theoretical. Evidently when it comes to practical matters, complications arise which may not be so easy to overcome. Moreover in real experimental situations, we do have much partial information about the degree of misalignment of the two frames which can be remedied by other possibly classical means. We briefly discuss these issues at the end.  Nevertheless it is worth to study this scheme in view of its generality (as we will see) and mathematical beauty when expressed for general groups. \\

The structure of this paper is as follows: In section (\ref{The QKD scheme}), we introduce a four state protocol which is meant to be the SRF-free version of the BB84 protocol and in section (\ref{six-state}) we present the SRF-free version of the 6-state protocol.  We end the paper with a discussion in which we also discuss generalizations to other groups or other reference frames and also briefly discuss some practical and experimental issues. 

\section{A  four-state  protocol without shared reference frame}\label{The QKD scheme}
To suggest an SRF-free analog of the BB84 protocol, 
Alice and Bob need a two dimensional space and two different bases for preparation and measurements of states in this space. The states should not have any reference to any coordinate axis of the two players, since such states are not invariant under the map (\ref{basicc}).  
  If we take two spin $1/2$ particles, their total spin can be $0$ or $1$. At first glance, it seems that we can take these two states to define our Hilbert space. However  a 
super-selection rule allows superposition of only those basis states which have the same total spin. More  generally and concretely, suppose that Alice and Bob have frames which are connected by a group of transformations $G$ and let Alice prepares a state $|\psi\ra_A=a|\m\ra+b|\n\ra$ where $|\m\ra$ and $|\n\ra$ are two states which transform under two inequivalent irreducible representations of the group $G$.  We now use  (\ref{basicc}),  and the Shur's two lemmas according to which if for a matrix $M$,  $U_\m(h)M{U_\n}^\dagger(h) =M\ \ \ \forall \ h\in G$, then $M\propto I$ for $\m=\n$ and $M=0$ for $\m$ inequivalent to  $\n$.  This immediately leads to the following state for Bob
\be
\rho_B=|a|^2|\m\ra\la\mu|+|b|^2|\n\ra\la \n|
\ee
which has completely lost its original coherence. \\

\begin{figure} 
	\includegraphics[width=16cm,height=12cm]{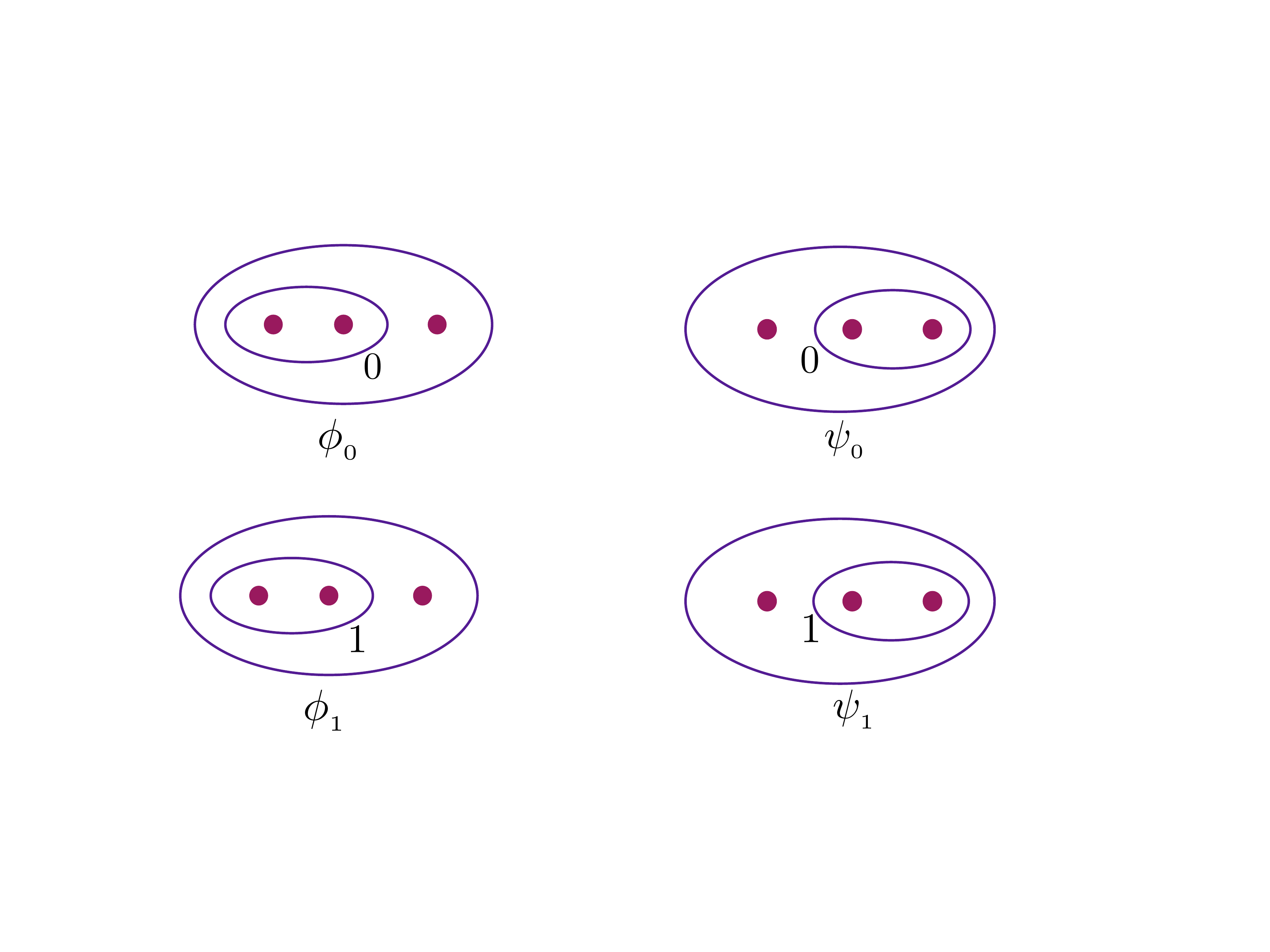}
	\vspace{-3cm}\caption{ Color online: Two different bases for the subspace of three particles with total spin 1/2, Eqs. (\ref{phiphi}) and (\ref{psipsi}). Alice randomly encodes her bits into one of the these two bases, Bob also measures in one of these bases. }
	\label{state}
\end{figure}

The simplest solution is to take two states of three particles both of which have the same total spin, but have different internal spins. That is we consider the two basis states of a qubit to be the two states which have total spin equal to $\frac{1}{2}$, while the  spin of the pair $(1,2)$ are $0$ and $1$, figure (\ref{state}). They can be written as 

\begin{equation}\label{set1}
|\phi^m_0\rangle := |\dfrac{1}{2}^{123} , 0^{12} , m\rangle,\h
|\phi^m_1\rangle := |\dfrac{1}{2}^{123} , 1^{12} , m\rangle,
\end{equation}
where $m=\pm$ is the z-component of the total spin.  
This is reminiscent of what we have in topological quantum computation, when the Hilbert space of anyons corresponds to the various ways that a specific number of anyons can fuse to get a  total charge with a specific value \cite{kitaev}.  For the same space, one can choose another basis, denoted by $|\psi_0\ra$ and $|\psi_1\ra$, where this time, the  spin of the last two particles, namely the pair $(2,3)$ is either $0$ or $1$, figure (\ref{state}). 
\begin{equation}\label{set2}
|\psi^m_0\rangle := |\dfrac{1}{2}^{123} , 0^{23} , m\rangle,\h
|\psi^m_1\rangle := |\dfrac{1}{2}^{123} , 1^{23} , m\rangle,
\end{equation}

The explicit form of these states are 

\begin{equation}\label{phiphi}
|\phi^+_0\rangle := \frac{1}{\sqrt{2}}(|+,-,+\rangle-|-,+,+\ra),\h
|\phi^+_1\rangle := \frac{1}{\sqrt{6}}(|+,-,+\ra+|-,+,+\ra-2|+,+,-\ra),
\end{equation}

and

\begin{equation}\label{psipsi}
|\psi^+_0\rangle := \frac{1}{\sqrt{2}}(|+,+,-\rangle-|+,-,+\ra),\h
|\psi^+_1\rangle := \frac{1}{\sqrt{6}}(|+,+,-\ra+|+,-,+\ra-2|-,+,+\ra),
\end{equation}

with $|\phi_i^-\ra$ and $|\psi_i^-\ra$ obtained from the above states by flipping all the spins.  \\

%Note that $|\phi_0\ra$ and $|\phi_1\ra$ are respectively anti-symmetric and symmetric with respect to the first two particles. This makes them orthogonal to each other and also perfectly distinguishable by an appropriate measurement. The same type of symmetry exists for states $|\psi_0\ra$ and $|\psi_1\ra$, with respect to interchange of particles $2$ and $3$. \\

The states have the following inner products:

\be \label{inner product}
\la \phi^{\pm}_i|\psi^{\pm}_j\ra=S_{i,j},\h \la \phi^{\pm}_i|\psi^{\mp}_j\ra=0, 
\ee
where $S$ is the following matrix

\be \label{S-matrix}
S=\frac{1}{2}\left(\begin{array}{cc} 1&\sqrt{3}\\\sqrt{3}&-1\end{array}\right).
\ee

Suppose now that Alice sends the states $|\phi_0^+\ra$ and $|\phi_0^-\ra$ with equal probability to Bob.  The statistics of measurements by Bob is, as if, Alice is sending him the state 

\be \label{sent state 1}
\rho_0:=\frac{1}{2}\left(|\phi_0^+\ra\la\phi_0^+ |+|\phi_0^-\ra\la\phi_0^- |\right).
\ee
This state being the projector to spin $0$ of the pair (1,2), is invariant under the channel ${\cal E}$ and when Bob measures the total spin of the pair (1,2), he obtains with probability 1, the spin 0. 
The other states sent by Alice are as follows:

\ba \label{sent state 2}
\rho_1&=&\frac{1}{2}\left(|\phi^+_1\ra\la \phi^+_1 |+|\phi^-_1\ra\la\phi^-_1 |\right),\cr
\sigma_0&=&\frac{1}{2}\left(|\psi^+_0\ra\la \psi^+_0 |+|\psi^-_0\ra\la\psi^-_0 |\right),\cr
\sigma_1&=&\frac{1}{2}\left(|\psi^+_1\ra\la \psi^+_1 |+|\psi^-_1\ra\la\psi^-_1 |\right).
\ea

{\bf Remark:}{\it {The states that Alice sends are always pure and of the form $|\phi_i^{\pm}\ra$ (with equal probability) or $|\psi_i^{\pm}\ra$ (with equal probability). The statistics produced corresponds to the above mixed states.  } }\\

The states $\rho_i$ have total spin $S=i$ for the pair $(1,2)$ and the states $\sigma_i$ have total spin $S=i$ for the pair $(2,3)$. From (\ref{inner product}), one can infer the inner product of these states:

\be \label{matrix inner product}
Tr(\rho_i\rho_j)=Tr(\sigma_i\sigma_j)=\frac{1}{2}\delta_{i,j}\h Tr(\rho_i\sigma_j)=\frac{1}{2}(S_{i,j})^2.
\ee

 It is also important to note that any rotation in the coordinate system of Alice, does not change the total values of spins for the three particles or the pairs (1,2) and (2,3) and henceforth the symmetry or antisymmetry of these states with respect to the aforementioned interchanges. More precisely, suppose that the frames of Bob and Alice are not perfectly aligned and we have only partial information about their alignment in the form of a probability distribution $P(R)$, where $R\in SO(3)$ is the rotation necessary to align them. Then any state $\rho$ which is sent by Alice,  seems to go through a channel 
 
\be\label{rhobob}\rho_A\lo \rho_B=\int dR P(R)U(R)^{\otimes 3}\rho_A U^{\dagger}(R)^{\otimes 3},\ee
and is received by Bob. The point is that the states given in (\ref{sent state 1}) and (\ref{sent state 2}), are all invariant states of this quantum map. \\

%Therefore when Alice sends a state $|\phi^+_i\ra$, the state received by Bob no longer has a fixed total spin in the $z$ direction, but still has the same total spin ($S^{1,2,3}_{tot}$) for the three particles $(1,2,3)$ and the same total spin $S^{1,2}_{tot}$ for the particles $(1,2)$. This state can be written as 
 
 \iffalse
 \be
 {\cal E}(|\phi^+_i\ra \la \phi^+_i|)=p_+|\phi^+_i\ra\la \phi^+_i|+ p_+|\phi^-_i\ra\la \phi^-_i|,  \ee
  \noindent where $p_{\pm}$ are determined by the probability distribution $P(R)$ and are independent of the sent state. \\

Alice now encodes $0$ and $1$ randomly into one of the above two sets of states, i.e. she encodes $0$ either in $|\phi^+_0\ra$ or $|\psi^+_0\ra$ and $1$ either in $|\phi^+_1\ra$ or $|\psi^+_1\ra$ and sends the state to Bob.\\
\fi

Note that Bob, having no shared reference frame with Alice,  can only perform optimal measurements which are total spin projectors of either the pair (1,2) or the pair (2,3). For ease of notation and to make the scheme parallel to the BB84 protocol, we call these the two bases of measurements of Bob, and simply call them the $\rho$ basis and the $\sigma$ basis respectively. In other words, Bob randomly uses one of the following two sets of projective measurements:

\begin{equation}
\begin{aligned}
E_{\rho} =\lbrace (\Pi_0)_{12}, (\Pi_1)_{12} \rbrace\\
E_{\sigma}=\lbrace (\Pi_0)_{23},  (\Pi_1)_{23}\rbrace.
\end{aligned}
\end{equation}
 Interestingly, in view of equations (\ref{sent state 1}) and (\ref{sent state 2}),  these two projectors are proportional to the mixed states $\rho$ and $\sigma$. Therefore we have 

\begin{equation}\label{projectors}
\begin{aligned}
E_{\rho} =\lbrace 2\rho_0, 2\rho_1 \rbrace,\\
E_{\sigma}=\lbrace 2\sigma_0, 2\sigma_1 \rbrace.
\end{aligned}
\end{equation}
  
  Let us remind the reader of the basics of the protocol. Alice encodes the bit $i=0,1$ randomly into the mixed states $\rho_i$ or $\sigma_i$. In practice she sends pure states of the form $\phi_i^{\pm}$ or $\psi_i^\pm$ with equal probability. Bob randomly chooses one of the 
   POVM's $E_\rho$ and $E_\sigma$. Let us for simplicity call these bases both for preparation and measurements the $\rho$ and the $\sigma$ bases.\\
   
    Let $P_{\rho,\rho}(B=i|A=j)$  denote the probability that Bob obtains the value $i$ when Alice sends the bit value $j$, in case that both use the $\rho$ bases. With similar definition for similar expressions.  Then it is clear from (\ref{matrix inner product}) that 
\ba
P_{\rho,\rho}(B=i|A=j)&=&Tr((\Pi_{i})_{1,2}\rho_j)=2Tr(\rho_i\rho_j)=\delta_{i,j}\cr
P_{\sigma,\sigma}(B=i|A=j)&=&Tr((\Pi_{i})_{2,3}\sigma_j)=2Tr(\sigma_i\sigma_j)=\delta_{i,j}.
\ea
Therefore in those rounds where they use the same bases,  we have perfect correlation between the bits sent by Alice and measured by Bob. On the other hand, if they use different bases, then we will have
\ba
P_{\rho,\sigma}(B=i|A=j)&=&Tr((\Pi_{i})_{2,3}\rho_j)=2Tr(\sigma_i\rho_j)=(S_{i,j})^2\cr
P_{\sigma,\rho}(B=i|A=j)&=&Tr((\Pi_{i})_{1,2}\sigma_j)=2Tr(\rho_i\sigma_j)=(S_{i,j})^2.
\ea
In those rounds where the bases are not the same, perfect correlation is lost between the bits, and these rounds are discarded after public announcement of the bases by the two parties.
 The important point is that the basic concept and methodology of the BB84 protocol is also at work here, i.e. random preparation and measurements in two bases, public announcement of bases and discarding those rounds where the bases do not match.\\
 
 {\bf Remark:}{\it  Furthermore note that Alice and Bob can randomly rotate their coordinate systems, without affecting the performance of the protocol. In this way, they prevent Eve from aligning her coordinate system with those of Alice or Bob and possibly doing non-invariant measurements on single particles which leak information to her.}\\

To generalize this method to those cases where there is no common frame of reference (be it a phase or a coordinate reference or any other reference pertaining to a group G), one should encode the bits in the internal charge of composite systems, where the states of these composite systems are themselves invariant under the twirling operation of the group (for example equation (\ref{rhobob}) for the rotation group). In other words the states which encode the bits are the basis states of the fusion space of the three particles. This internal charge (in our case, spin) is hidden from the adversaries who do not know in which basis to measure the charge. 
In this way, Alice and Bob establish a shared random key between themselves which turns out to be secure against attacks by Eve in the same way that BB84 is secure \cite{BB84-security-1,BB84-security-2,BB84-security-3}, as long as the resources available to Eve are strong enough to manipulate the package of three particles.  It should be noted that any adversary is prohibited from assessing bits of the key, since she cannot align her reference frame with both frames in possessions of the two parties. In fact  Alice and Bob can even randomly rotate their frames in different rounds without affecting the performance of the protocol. In the absence of such a common frame, the only measurement which leads to meaningful results for Eve is the total spin measurement of pairs of spins. In the same way as in the BB84 protocol, she has to measure the total spins of the pairs in random. In this way,	in half of those rounds where the so called bases of Alice and Bob do agree, Eve chooses the wrong basis for measurement. It is then seen that in total  she incurs error on bits shared between Alice and Bob at a rate of $3/16$. To see this, consider as an example the case where Alice and Bob both choose the basis $\rho$ but Alice sends $0$ and  Bob receives $1$ due to Eve's intervention. The probability of this is given by 
		\ba \label{error rate}
	P_{\rho,\rho}^{error}(B=1|A=0)&=&\frac{1}{2}	P_{\rho\sigma}(B=1|E=0)P_{\sigma\rho}(E=0|A=0)\cr &+&\frac{1}{2}P_{\rho\sigma}(B=1|E=1)P_{\sigma\rho}(E=1|A=0)\cr
	&=& \frac{1}{2}\left(|S_{10}|^2|S_{00}|^2+|S_{11}|^2|S_{10}|^2\right)=\frac{3}{16}.
	\ea
A similar analysis leads to the same value for the other cases, when the two parties are using the basis $\sigma$. Thus in this protocol, the intervention of Eve introduces an error rate of $\frac{3}{16}$, which can lead to detection of an adversary, when the two parties compare only a subsequence of the bits. \\	

{\bf Remark:}{\it Using tools from information theory, the absolute security of the BB84 protocol has been proved in \cite{BB84-security-1,BB84-security-2,BB84-security-3}. Our emphasis here is on the theoretical possibility of an SRF-free version of this protocol. Therefore  as far as information theory is concerned and even if we assume strong enough resources for Eve for manipulation of packages of three-particles, the scheme presented here is also absolutely secure. Of course practical implementation of this protocol is a different problem which naturally has its own complications like preparation and measurements of entangled states, lower bit rate by a factor of 3, particle losses, etc.  }\\

\section{A six state protocol without shared reference frame}\label{six-state}
Having three different bases, it is only natural to consider the analog of the six-state protocol \cite{6-state}, by including the third pair of spins, namely the pair $(1,3)$ and encode the qubit in the charge of this pair.
The protocol is based on the use of three sets of preparation and measurements rather than two sets. That is, Alice uses the total spin of $(1,2)$, $(2,3)$ or $(1,3)$ for encoding and Bob correspondingly have three sets of POVM's.   
More precisely, in addition to the two sets of state given in (\ref{set1}) and (\ref{set2}) Alice can also encode her bits $0$ and $1$ into the following set of states, whose spin of the pair $(1,3)$ is $0$ or $1$. 
\begin{equation}
|\chi^+_0\rangle := \frac{1}{\sqrt{2}}(|+,+,-\rangle-|-,+,+\ra),\h 
|\chi^+_1\rangle := \frac{1}{\sqrt{6}}(|+,+,-\ra+|-,+,+\ra-2|+,-,+\ra).
\end{equation}
As in the previous cases, Alice encodes a bit $i$ with equal probability into $|\chi^+_i\ra$ and $|\chi_i^-\ra$.
The new states by Alice are described by the following density matrices 
\ba
\gamma_0&=&\frac{1}{2}\left(|\chi^+_0\ra\la \chi^+_0|+|\chi^-_0\ra\la\chi^-_0 |\right),\cr
\gamma_1&=&\frac{1}{2}\left(|\chi^+_1\ra\la \chi^+_1|+|\chi^-_1\ra\la\chi^-_1|\right),
\ea
and the new POVM used by Bob (measuring the total spin of the pair (1,3)) is given by
\begin{equation}\label{projectors3}
E_{\gamma} =\lbrace 2\gamma_0, 2\gamma_1 \rbrace,\\
\end{equation}

Alongside equation (\ref{inner product}), we also have the following inner products:

\be
\la \phi^\pm_i|\chi^\pm_j\ra=\la \psi^\pm_i|\chi^\pm_j\ra=S_{ij},
\ee
with the same matrix as in (\ref{S-matrix}).  This leads to the following relation in addition to (\ref{matrix inner product})
\be
Tr(\gamma_i\gamma_j)=\frac{1}{2}\delta_{i,j}\h Tr(\gamma_i\rho_j)=Tr(\gamma_i\sigma_j)=\frac{1}{2}(S_{i,j})^2.
\ee

The percentage of valid rounds (e.g. the rounds which are not discarded) now drops from $1/2$ to $1/3$, but the error rate induced by intervention of an adversary  increases from $\frac{3}{16}$ to $\frac{4}{16}=\frac{1}{4}$. This is simply seen by following the same steps that led to Eq (\ref{error rate}). \\

\begin{figure} 
	\includegraphics[width=14cm,height=10cm]{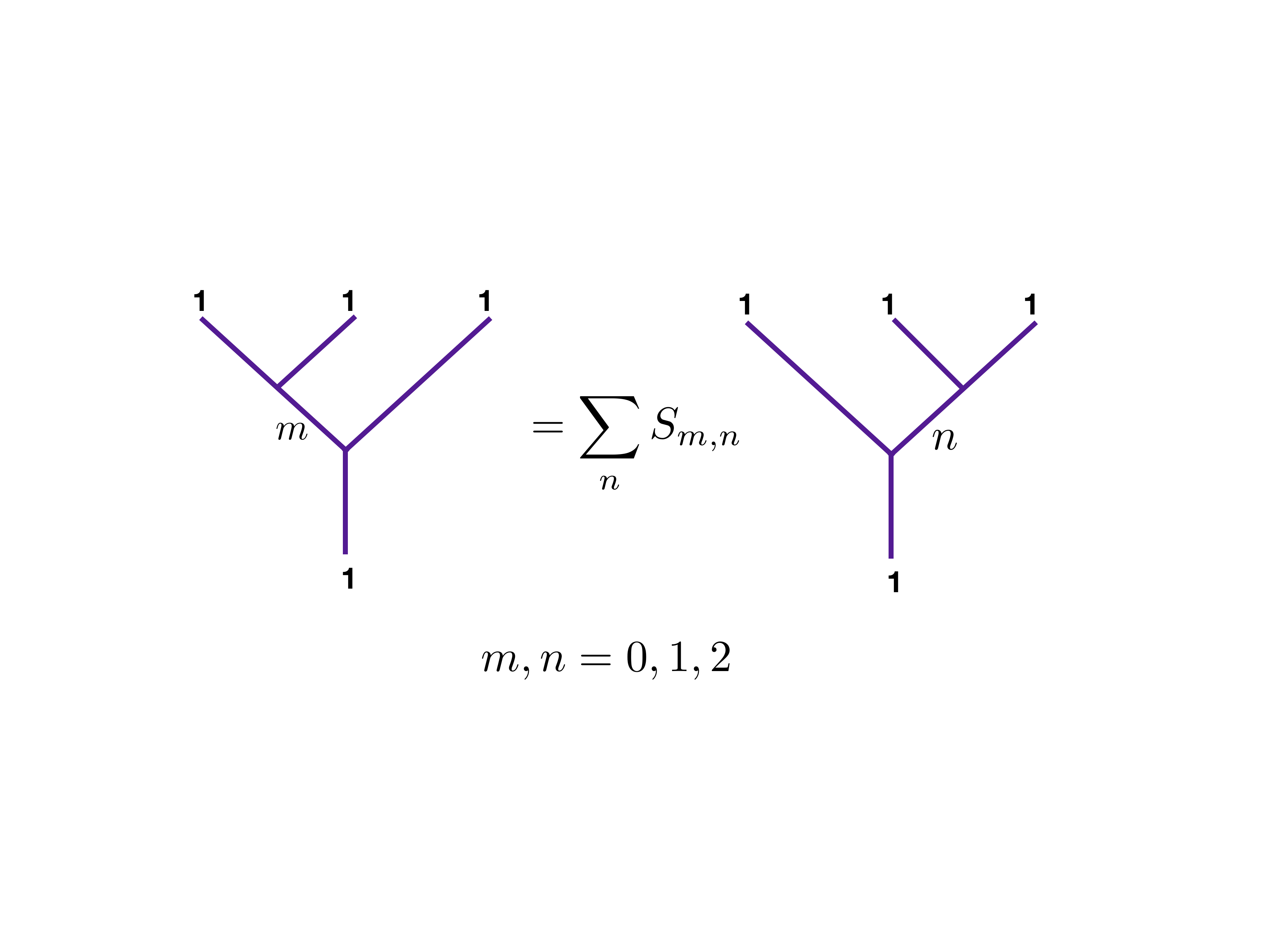}
	\vspace{-2.5cm}\caption{ Color online:  } The fusion space of three spin 1 particles, can be used to encode the values 0, 1 and 2 and perform an SRF-free version of the BB84 QKD system with qutrits. In principle the scheme can be generalized to every other group. 
	\label{states2}
\end{figure}

All these considerations can be generalized to other representations of $SU(2)$ and other non-abelian groups. For example if the particles have spin 1, then we have the decomposition rule  

\be
1\otimes 1\otimes 1=0\oplus 1^3\oplus 2^2\oplus 3,
\ee
where the superscripts show the multiplicities of these representations. It shows for example that there are three different paths of fusion all leading to the total spin 1. Each path  corresponds to a state with specific internal spins (charge) for the pair of particles (1,2), Fig. \ref{states2}. Thus in the same way that we have shown above, these states and the pair of particles which are chosen for measurements can act as a QKD system with three-level states (qutrits) in the complete absence of reference frames. The same method can be used if instead of the rotation group, $G$ is any other non-abelian group. It is true that in practical situations, there are  more feasible ways, either by actively  aligning the coordinate systems \cite{active alignment QKD1,active alignment QKD2} or by using other degrees of freedom, like angular momentum of light \cite{RFI-QKD3}. It is also true that we may have partial information about the two coordinate systems, like one common direction \cite{RFI-QKD1,RFI-QKD2,RFI-QKD4}, which alleviates the need for these schemes. Nevertheless exploring the theoretical possibility of achieving such general schemes for any group of transformation is interesting. Finally with advances in control and manipulation of entangled states of photons\cite{pan}, the scheme proposed in this paper may also find practical applications.  

\section{conclusion} \label{conclusion}
We have discussed QKD protocol in the absence of shared reference frame from a general point of view, where the two frames are related to each other by elements of a general group $G$. To define the versions of BB84 or 6-state protocols for a general group $G$, we need to define reference-frame-free states and measurements. In the minimal scheme, three particles should be used  and the states are those who have identical total charge but different patterns of internal charges  for creating that total charge. In the terminology used for anyons  in the context of topological quantum computation \cite{kitaev}, these different patterns are called fusion paths Fig. (\ref{states2}). For each total charge, the number of different paths determine the dimension of the Hilbert space, i.e. QKD with qubits or qudits.  Where all these considerations apply to any group relating two different reference frames, we have used the case of the coordinate reference frame and the rotation group for definiteness and probably for its  possible  practical relevance, i.e. in those situations where the two stations are rotating or where noises in optical fibers or turbulence in free space communication is too high.  Of course we do not emphasize this practical aspect since, needless to to say,  implementation of these protocol is a different problem which naturally has its own complications like preparation and measurements of entangled states, lower bit rate by a factor of 3, particle losses, etc. Furthermore, in practical situations, there may be more feasible ways, for active alignment of coordinate systems \cite{rmk1,rmk2, beheshti}, or use other more specific protocol for coordinate reference frames \cite{berg}. \\

{\bf Acknowledgments:} The authors thank  Marzieh Bathaee and Sadegh Raeisi for valuable comments. Their special thanks also go to Mauro Paternostro for a series of very instructive discussions through email. This work was partially supported by the  grant No. G950222 from the vice-chancellor of Sharif University of Technology. The works of A. Mani and F. Rezazadeh was supported by a grant no. 96011347 from Iran National Science Foundation. Vahid Karimipour also thanks Abdus Salam ICTP where the final stages of this work was done and the Simons Foundation for financial support through the ICTP associate program.

{}


\begin{thebibliography}{}
	
\bibitem{aligning reference frame1}	E. Bagan, M. Baig, and R. Mu˜noz-Tapia, Phys. Rev.Lett. \textbf{87}, 257903 (2001).
\bibitem{aligning reference frame2}  G. Chiribella, G. M. D’Ariano, P. Perinotti, and M. F. Sacchi Phys. Rev. Lett. \textbf{93}, 180503 (2004).
\bibitem{aligning reference frame3} A. Peres and P. F. Scudo, Phys. Rev. Lett. \textbf{86}, 4160 (2001).
\bibitem{aligning reference frame4} A. Peres and P. F. Scudo, Phys. Rev. Lett. \textbf{87}, 167901 (2001).
\bibitem{communication without SRF1} L. Aolita and S. Walborn, Phys. Rev. Lett. \textbf{98}, 100501 (2007).
\bibitem{communication without SRF2} S.D. Bartlett, T. Rudolph, and R.W. Spekkens, Phys.Rev. Lett. \textbf{91}, 027901 (2003).
\bibitem{marvian} G. Gour, I. Marvian, and R.W. Spekkens, Phys. Rev. A \textbf{80}, 012307 (2009).

\bibitem{nonlocality without SRF1}  C. F. Senel, T. Lawson, M. Kaplan, D. Markham and E. Diamanti, Phys. Rev. A \textbf{91}, 052118 (2015).
\bibitem{nonlocality without SRF2} J. J. Wallman and S. D. Bartlett, Phy. Rev. A \textbf{85}, 024101 (2012).
\bibitem{nonlocality without SRF3}  F. Verstraete and J. I. Cirac, Phys. Rev. Lett. \textbf{91}, 10404 (2003).

\bibitem{RFI-QKD1} A. Laing, V. Scarani, J. G. Rarity, and J. L. O’Brien, Phys.
Rev. A \textbf{82}, 012304 (2010).
\bibitem{RFI-QKD2}  C. E. R. Souza, C. V. S. Borges, A. Z. Khoury, J. A. O. Huguenin, L. Aolita, and S. P. Walborn, Phys. Rev. A \textbf{77}, 032345 (2008).
\bibitem{RFI-QKD4} J.A. Slater, C. Branciard, N. Brunner, and W. Tittel, New J. Phys. \textbf{16}, 043002 (2014).

	
\bibitem{BB84} C. H. Bennett and G. Brassard, in Proc. IEEE International Conference on Computers, Systems, and Signal Processing, Bangalore, India, 1984 (IEEE, New York, 1984), p. 175.
\bibitem{Ekert} Ekert. A, Phys. Rev. Lett. \textbf{67}, 661-663 (1991).
\bibitem{6-state} D. Bruss. Phys. Rev. Lett. \textbf{81}, 3018 (1998).
\bibitem{cerf}  Nicolas J. Cerf, Mohamed Bourennane, Anders Karlsson, and Nicolas Gisin, Phys. Rev. Lett. 88, 127902 (2002).
\bibitem{karl} M. Bourennane, A. Karlsson, and G. Bjo\ ̈rk, Phys. Rev. A \textbf{64}, 052313 (2001).
\bibitem{vk} V Karimipour, A Bahraminasab, S Bagherinezhad, Phys. Rev. A \textbf{65} 052331 (2002).


\bibitem{SSR1} G. Gour and R.W. Spekkens, New J. Phys. \textbf{10}, 033023 (2008).
\bibitem{SSR2} S.D. Bartlett, T. Rudolph, and R.W. Spekkens, Rev. Mod. Phys. \textbf{79}(2):555-609 (2007).
\bibitem{kitaev} M. H. Freedman, A. Yu. Kitaev, and Z. Wang, Commun. Math. Phys. 227 587-603, (2002); M. Freedman, M. Larsen, and Z. Wang, Comm.Math. Phys. 227 605622, (2002); M. H. Freedman, A. Kitaev, M. J. Larsen, and Z. Wang, Math. Soc. 40 31-38, (2003).
\bibitem{BB84-security-1} H. K. Lo, H. F. Chau, Science \textbf{283} 2050-2056 (1999).
\bibitem{BB84-security-2} D. Mayers, JACM \textbf{48}, no 3, 351-406 (2001).
\bibitem{BB84-security-3} E. Biham, M. Boyer, P. Oscar Boykin, T. Mor, V. Roychowdhury, Journal of Cryptology archive \textbf{19}, Issue 4, 381-439 (1999).


\bibitem{active alignment QKD1} H.L. Yin,T.Y. Chen,Z.W. Yu,H. Liu,L.X. You, Y.H. Zhou, S.J. Chen, Y. Mao, M.Q. Huang, W.J. Zhang, H. Chen, M.J. Li,D. Nolan, F. Zhou, X. Jiang, Z. Wang, Q. Zhang, X.B. Wang, J.W. Pan, Phys. Rev. Lett, \textbf{117}, 190501 (2016). 
\bibitem{active alignment QKD2}  L. C. Comandar et al, Nature Photon, vol. 10 ,no. 5, pp. 312–316, (2016).

\bibitem{RFI-QKD3} F. M. Spedalieri, Optics Communications \textbf{260}, 340 (2006).  
\bibitem{preskill}
\bibitem{rmk1} F. Rezazadeh, A. Mani, and V. Karimipour, "Secure alignment of coordinate systems by using quantum correlation",  Phys. Rev. A \textbf{96}, 022310 (2017).
\bibitem{rmk2} F. Rezazadeh, A. Mani, and V. Karimipour, "Shared entangled states as a substitute for shared reference frames", preprint: arXiv:1811.05657. 
\bibitem{beheshti} A. Beheshti, S. Raeisi, and V. Karimipour, "Entanglement-assisted communication in the absence of shared reference frame", preprint, arXiv:1901.01503, to appear in Phys. Rev. A.
\bibitem{pan} Anton Zeilinger, 
Physica Scripta, Volume 92, Number 7 (2017).
\bibitem{berg} G. Tabia,  and B-G.  Englert, Physics Letters A. \textbf{375} 817-822 (2011).

\end{thebibliography}
\end{document}